\documentclass[aps,showpacs,twocolumn]{revtex4}

\usepackage{bm}
\usepackage{epsfig}
\usepackage{graphicx}

\begin{document}

\title{Probing CP violation with the electric dipole moment of atomic mercury}

\author{K. V. P. Latha $^{\star}$, D. Angom $^{\dag}$,  
        B. P. Das$^{\star}$ and  D. Mukherjee $^{\star \star}$ } 
     \affiliation{ 
         ${\star}$ Indian Institute of Astrophysics, Bangalore, India.}  
     \affiliation{ 
         $\dag$ Physical Research Laboratory, Ahmedabad, India.} 
      \affiliation{ 
         $^{\star \star}$  Indian Association of Cultivation of Science, 
                           Kolkata, India. \\ 
       }


\begin{abstract}
   The electric dipole moment of atomic $^{199}$Hg induced by the nuclear
Schiff moment and tensor-pseudotensor electron-nucleus interactions has been
calculated. For this, we have developed and employed a novel method based on 
the relativistic coupled-cluster theory. The results of our theoretical 
calculations combined with the latest experimental result of $^{199}$Hg 
electric dipole moment, provide new bounds on the T reversal or CP violation 
parameters $\theta_{\rm QCD}$, the tensor-pseudotensor coupling constant 
$C_T$ and $(\widetilde{d}_u - \widetilde{d}_d)$.  This is the most accurate 
calculation of these parameters to date. We highlight the the crucial role 
of electron correlation effects in their interplay with the P,T violating 
interactions. Our results demonstrate substantial changes in the results of 
earlier calculations of these parameters which can be attributed to the more 
accurate inclusion of important correlation effects in the present work.
\end{abstract}



\maketitle


   The existence of a nonzero permanent electric dipole moment of a 
non-degenerate physical system is a signature of the simultaneous violations 
of parity (P) and time-reversal (T) symmetries \cite{landau}. T violation 
implies the combined charge conjugation (C) and P violation ; i.e., CP 
violation via the CPT theorem \cite{luders}. The origin of CP violation is 
still not well understood. It has so far been observed only in the $K$ 
\cite{christenson-64} and $B$ \cite{abe-01,aubert-01} mesons and the results 
are essentially in agreement with the predictions of Kobayashi and Maskawa 
in the framework of the Standard Model \cite{ibrahim}. However, this model 
cannot explain the matter-antimatter asymmetry \cite{dine-04} in the 
universe for which CP violation is a pre-requisite \cite{gavela}. In 
addition, it predicts atomic EDMs several orders of magnitude below their 
current limits \cite{pospelov,barr1993}. Indeed, atomic EDMs are excellent 
probes of physics beyond the Standard Model \cite{pospelov,barr1993} and 
they provide important insights into a rich variety of CP violations--leptonic,
semi-leptonic and hadronic sectors. Experimental searches are underway for 
the EDMs of para-magnetic (open-shell) and dia-magnetic (closed-shell) atoms 
\cite{exp-para,griffith-09}. The results of the experiments can be combined 
with those of sophisticated atomic many-body calculations to determine 
various CP violating coupling constants at the atomic level which can 
ultimately be related to the CP violation parameters at the elementary 
particle level \cite{barr1992}.  The EDM of dia-magnetic atoms arises 
predominantly from the nuclear Schiff moment (NSM) and/or the electron-nucleon 
tensor-pseudotensor interactions \cite{barr1992}. These in turn arise from 
the nucleon-nucleon interactions or the EDM of nucleons, which originate due 
to the quark-quark interactions, EDMs and chromo EDMs of quarks at the 
elementary particle level. 

In the present work, we concentrate only on the EDM of mercury ($^{199}$Hg),
a closed-shell atom. The present limits on important CP violation parameters 
like, $\theta_{\rm QCD}$ for strong interactions and the chromo EDMs of
quarks have been obtained from the EDM of $^{199}$Hg \cite{griffith-09}, which
is the most accurate of all the data from atomic EDMs to date. The focus of 
our work in this Letter is to improve the current limits on the coupling 
constants associated with the electron-nucleon tensor-pseudotensor (T-PT) 
interaction ($C_T$) and the NSM($S$). A nonzero value of $C_T$ implies 
physics beyond the Standard Model. The dependence of the T-PT interactions 
and the NSM on the nuclear spin makes closed-shell atoms, in particular, 
those having nonzero nuclear spin the best candidates to measure EDMs 
sensitive to the nuclear sector.


For heavy atoms like mercury, it is customary to use the Dirac-Coulomb 
Hamiltonian $H_{\rm DC}$, in atomic units
\begin{equation}
H_{\rm DC}  =  \sum_i \left [ c\bm{\alpha}_i\cdot\bm{p}_i + 
                     \bm{\beta}_i mc^2 + V_N(r_i) \right ] + 
                     \sum_{i<j} \frac{1}{|\bm{r}_i -\bm{r}_j|} ,
  \label{h_dc}
\end{equation}
where $r_i$ refers to the electron coordinates, $\alpha$ and $\beta$, the Dirac
matrices and $V_N(r_i)$, the nuclear potential, and the last term is the 
electron-electron Coulomb interaction. The key and the most challenging step 
in atomic many-body physics is to incorporate the effects of 
electron-electron Coulomb interaction, last term in Eq.(\ref{h_dc}), as 
accurately as possible. Under the independent particle and the central field 
approximations \cite{lindgren}, with the introduction of the Dirac-Fock 
potential, the above Hamiltonian can be separated into an exactly 
solvable part ($H_0$) and a residual interaction part which consists of the 
Coulomb interaction and the Dirac-Fock potential \cite{lindgren}. The single 
particle wave functions are computed self-consistently from $H_0$ and the 
many particle wavefunctions are expressed as Slater determinants built out 
of the single particle wavefunctions. The effects of the residual Coulomb 
interaction are calculated with many-body methods. For this, finite order 
many-body perturbation theory (MBPT) and the configuration interaction (CI) 
approach are two widely used methods \cite{lindgren}. An even superior method, 
with strong theoretical many-body physics underpinnings is the 
coupled-cluster theory. In the present work, the P and T violating 
interactions  are treated perturbatively to first order in addition to the 
electron-electron residual Coulomb interaction within the framework of the 
relativistic coupled-cluster theory.

  The wavefunction in coupled-cluster theory has an exponential structure, 
see \cite{bartlett-07} for a recent review of the method, to describe 
correlation effects in many-body systems. It is non-perturbative and defines 
atomic states as superposition of states of different levels of excitations, 
which arise from the residual Coulomb interaction, with respect to a reference 
state.   

Mathematically, the coupled-cluster wavefunction can be expressed as  
\begin{equation}
  |\Psi_i\rangle = e^{T^{(0)}}|\Phi_i\rangle ,
\end{equation}
where $|\Phi_i\rangle $ is the reference state containing a fixed number
of electrons and $T^{(0)}$ is an operator which excites electrons out of it, 
thereby giving rise to states with different levels of excitations 
corresponding to different many-body effects. In our calculations we
use the coupled-cluster singles and doubles approximation, that is 
$T^{(0)} = T^{(0)}_1 + T^{(0)}_2$. In second quantized form
$$
T^{(0)}_1=\sum_{a,p}a_p^\dagger a_a t_a^p(0) \text{  and  }
T^{(0)}_2=\sum_{a,b,p,q}a_p^\dagger a_q^\dagger a_b a_a t_{ab}^{pq}(0),
$$ 
excite one and two electrons respectively from the reference state.  The 
equations that determine the amplitudes of $T^{(0)}$ are a set of coupled 
non-linear algebraic equations and these are solved iteratively till 
convergence.


  For closed-shell atoms, as mentioned earlier, one prominent source of EDMs 
is the nuclear Schiff moment $\bm{S}$ (NSM), a P and T odd electromagnetic 
moment of the nucleus. For a finite size nucleus of radius $R_N$, the Schiff 
moment potential \cite{flambaum-02} is 
\begin{equation}
  \varphi(\bm{R}) = -\frac{3\bm{S}\cdot\bm{R}}{B}\rho(R),
\end{equation}
where $B=\int \rho(R)R^4dR$ and $\rho(R)$ is the nuclear density. This 
potential interacts electrostatically with the electrons, it mixes atomic 
states of opposite parities and generates a finite atomic EDM, $d_A$. Then, 
the atomic Hamiltonian is $ H_{\rm atom} = H_{\rm DC} + \lambda H_{\rm PTV}$, 
where $H_{\rm PTV}^{\rm Schiff}=- \varphi(\bm{R})$ is the P and T  
violating interaction Hamiltonian and $\lambda$ is a T or CP violation 
parameter which can be considered as perturbation parameter. The eigenstates 
of the $H_{\rm atom} $ are the mixed parity states $|\widetilde{\Psi}\rangle$. 
To incorporate $H_{\rm PTV}$ as a first order perturbation, the exponential 
operator in coupled-cluster theory is redefined as 
$e^{T^{(0)}+\lambda T^{(1)}} $. The cluster operator $T^{(1)}$ has one order 
of $H_{\rm PTV}$ and mixes the states of opposite parities. As a result of 
this, the ground state
\begin{equation}
  |\widetilde{\Psi}_0\rangle= e^{T^{(0)}+\lambda T^{(1)}}|\Phi_0\rangle.
\end{equation} 
Then, unlike the $T^{(0)}$ equations, since $H_{\rm PTV}$ is considered 
to first order only, the equations for the amplitudes of $T^{(1)}$ are 
a set of linear algebraic equations,
\begin{equation}
\langle \Phi_0^\prime|\left[\overline H_{\rm N},T^{(1)}\right]|\Phi_0\rangle=-
      \langle \Phi_0^\prime|\overline H_{\rm PTV}|\Phi_0\rangle
\end{equation}
$\overline O = {e^{T^{(0)}}}^\dagger O e^{T^{(0)}}$ where $O$ is a general 
operator,  $H_{\rm N}$ is the normal-ordered Hamiltonian and $|\Phi'_0\rangle$ 
are opposite parity Slater determinants. Further, like in the unperturbed 
cluster operators $T^{(0)}$, we use the approximation 
$T^{(1)}=T^{(1)}_1 + T^{(1)}_2$. Then, the atomic EDM of the ground state is
\begin{equation}
  d_A = \frac{\langle \widetilde{\Psi}_0|D|\widetilde{\Psi}_0\rangle}
        {\langle \widetilde{\Psi}_0|\widetilde{\Psi}_0\rangle},
  \label{d_a}
\end{equation}
where $D$ is the electric dipole operator. In the above expression, after
expanding in terms of the cluster operators $T^{(0)}$ and 
$T^{(1)}$, only the terms first order in $T^{(1)}$ contribute. Often, 
$d_{\rm A}$ is computed perturbatively with the sum over states approach, 
which necessitates a truncation after the first few intermediate states. On 
the contrary, our relativistic coupled-cluster scheme doesn't involve 
summing over states and subsumes all possible intermediate states within the 
chosen configuration space.


   Besides the NSM, the other possible source of EDM in closed 
shell atoms is the tensor-pseudotensor electron-nucleus interaction
\begin{equation}
   H_{\rm PTV}^{\rm T-PT}= \frac{iG_F C_T}{\sqrt2} \sum_i \bm{\sigma}_N\cdot 
                \bm{\gamma}_i \rho_N(r),
\end{equation}
where $G_F$ is fermi constant, $C_T$ is coupling constant, $\bm{\sigma}_N$ is 
nuclear spin and $\bm{\gamma}_i$ is Dirac matrix. It must be emphasized that, 
this form of interaction does not exist within the Standard Model of particle 
physics and $C_T$ is zero. However, there are models which predict such an 
interaction \cite{barr1992}.

To extract the T or CP violation parameters, the atomic theory calculations 
are combined with the experimental data. In this context it is appropriate to 
rewrite Eq.(\ref{d_a}) as
\begin{equation} 
  d_A =  \lambda \eta,
  \label{eta}
\end{equation} 
where $\eta$ is the atomic enhancement factor. As defined earlier, the 
constant $\lambda$ is a T or CP violation parameter considered as a 
perturbation parameter. It can for example be the nuclear Schiff moment 
$\bm{S}$ or the coupling constant $C_T$. A precision atomic many-body 
calculation, like the coupled-cluster calculation reported here, would 
provide the value for a particular $\eta$. Experimentally, the measured 
atomic EDM $d_{\rm A}$ is the sum total of contributions from all the P and 
T symmetry violating phenomena within the atom. A bound on 
$\lambda = d_{\rm A}/\eta$ is obtained by combining the results of atomic 
theory and experimental data.  Depending on the choice of the atom, it is 
possible to identify the dominant sources of T or CP violation and derive 
tighter bounds.


For the present set of calculations, we employ the even-tempered Gaussian basis 
set expansion \cite{mohanti,chaudhuri-99}. The orbital basis set 
consist of $(1-18)s$, $(2-18)p_{1/2,3/2}$, $(3-13)d_{3/2,5/2}$, 
$(4-11)f_{5/2,7/2}$, $(5-9)g_{7/2,9/2}$ and $(6-7)h_{9/2,11/2}$. This orbital
basis set is considered complete for the coupled perturbed Hartree-Fock (CPHF) 
calculations.  That is, further increase in the number of orbitals does not 
change the results. In addition, we compute the ground state dipole scalar 
polarizability for $^{199}$Hg. We obtain a value $33.294 e a_0^3$, where 
$a_0$ is the Bohr radius, which is in excellent agreement with its 
experimental value \cite{miller,rad,goebel}.

  To date, among the closed-shell atoms, $^{199}$Hg as mentioned earlier sets 
the standard for the most precise EDM results. In a recent paper 
\cite{griffith-09}, the new upper limit is reported as
\begin{equation}
   |d(^{199}\text{Hg})| < 3.1 \times 10^{-29} e \text{ cm (95\% C. L.)}.
\end{equation} 
Our atomic calculation based on the relativistic coupled-cluster theory gives
\begin{equation}
   d_A^{\rm Schiff}(^{199}\text{Hg}) = -5.07 \times 10^{-17} 
        \left ( \frac{S}{ e \text{ fm}^3}\right ) e \text{ cm}.
  \label{da_hg}
\end{equation}
This is the first ever relativistic coupled-cluster result for any atomic 
EDM calculation  arising from the NSM. Combining with the experimental 
result, the limit on the NSM is
\begin{equation}
    S(^{199}\text{Hg}) < 6.1 \times 10 ^{-13} e \text{ fm}^3 .
\end{equation}
There is a large change of 96\% from the result of coupled perturbed
Hartree-Fock calculation
\begin{equation}
   d_A^{\rm Schiff}(^{199}\text{Hg}) = 2.8 \times 10^{-17} 
        \left ( \frac{S}{ e \text{ fm}^3}\right ) e \text{ cm},
\end{equation}
reported earlier \cite{dzuba-02}, which is in excellent agreement with the 
result of a similar calculation but in the framework of the relativistic 
coupled-cluster theory \cite{latha-08}. The large change in the two results 
demonstrates the importance of electron correlation effects and their 
interplay with the $H_{\rm PTV}^{\rm Schiff}$ interaction in determining the 
magnitude of the NSM.

\begin{figure}
  \includegraphics[width=6cm]{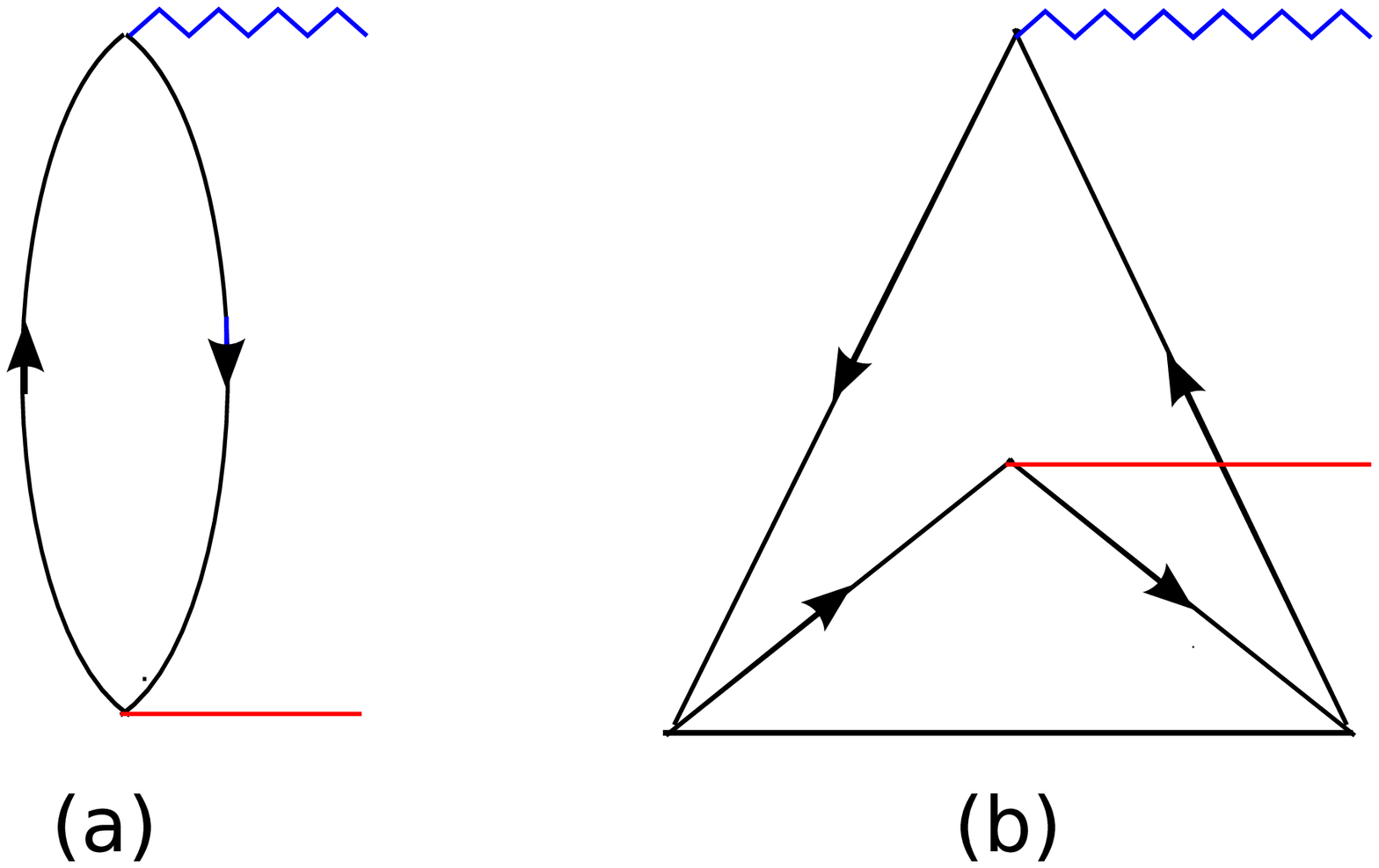}
  \caption{Two of the dominant many-body perturbation diagrams of  
           $d_{\rm A}$. Diagrams (a) and (b) represent 
           the terms ${T_1^{(1)}}^\dagger \overline D$ and 
           ${T_1^{(1)}}^\dagger D T_2^{(0)}$ respectively.
           At the interaction vertices, the horizontal wavy blue, black
           and  red lines represent the $H_{\rm PTV}$,  $T^{(0)}_2$ 
           and the electric dipole operator $D$ respectively. 
           }
  \label{diagedm}
\end{figure}

   It is possible to separate the contributions of individual terms in 
Eq.(\ref{d_a}) and the many-body perturbation diagrammatic representation of 
the dominant terms are shown in Fig.\ref{diagedm}. These diagrams represent 
the excitations and de-excitations due to cluster operators and dressed 
dipole operator $\overline D$. Earlier calculations 
\cite{dzuba-02,martensson} incorporate only a certain class of two-particle 
two-hole excitations which are subset of the correlation effects we have 
included through the cluster operator $T^{(0)}$ in the present calculation.
The two most dominant terms are ${T_1^{(1)}}^\dagger \overline D$  and 
${T_1^{(1)}}^\dagger DT_2^{(0)}$, in Eq.(\ref{da_hg}) these terms 
individually contribute 
$-5.40 \times 10^{-17} ( S/( e \text{ fm}^3)) e \text{ cm}$ and 
$-0.17 \times 10^{-17} (S/( e \text{ fm}^3) ) e \text{ cm}$ respectively.  
In Fig.\ref{diagedm}, the diagrammatic equivalent of these terms are (a) 
and (b) respectively.

 Our result of $^{199}$Hg EDM arising from the electron-nucleus 
tensor-pseudotensor interaction is 
\begin{equation}
  d_A^{\rm T-PT} = -4.3  \times 10^{-20}C_T\sigma_N e \text{ cm}.
\end{equation}
Compared to the CPHF result $-6.19 \times 10^{-20}C_T\sigma_N e$ cm 
\cite{martensson,latha-08}, the change with the additional correlation effects 
is not so dramatic.  There is a decrease of 31\%, which is significant 
but not so spectacular as in $d_A^{\rm Schiff}(^{199}\text{Hg})$. This 
comparison demonstrates, without any ambiguity, the importance of electron 
correlation effects in precision atomic EDM calculations. A closer examination 
of the structure of the two P and T violating Hamiltonians in the present 
work sheds some light on why the electron correlation effects are larger in 
the case of the NSM than the tensor-pseudotensor interaction. Both the 
$p_{1/2}$ and $p_{3/2}$ electrons are actively involved in the interplay of 
the $H_{\rm PTV}$ and electron correlation effects in the former, while the 
contribution of the $p_{3/2}$ electrons is negligible in the latter 
\cite{dzuba-02}. 

  The individual contributions follow similar trend as in the case of NSM.  
The terms ${T_1^{(1)}}^\dagger \overline D$ and 
${T^{(1)}}^\dagger \overline DT_2^{(0)}$ give the largest ($\approx $ 95\%) 
and the second largest contributions to 
$d_{\rm A}^{\rm T-PT}$, $-4.8  \times 10^{-20}C_T\sigma_N e \text{ cm}$ and 
$-0.27 \times 10^{-20}C_T\sigma_N e \text{ cm} $ respectively. Then, a limit
\begin{equation}
    C_T < 1.4 \times 10^{-9},
\end{equation}
is obtained after combining our results with the experimental data.

  Assuming that the NSM arises from the 
nucleon-nucleon interactions with pions as the dominant mediators
\cite{jesus-05}
\begin{eqnarray}
\bm{S}(^{199}\text{Hg}) & = & g_{\pi NN}\left[0.01\bar{g}_{\pi NN}^{(0)} + 0.07 
                     \bar{g}_{\pi NN}^{(1)} \right. \nonumber \\
                   &   & \left . + 0.02\bar{g}_{\pi NN}^{(2)}\right]
                         e\text{ fm}^3,
\end{eqnarray}
where $g_{\pi NN}$ and $\bar{g}_{\pi NN}^{(i)}$ are the CP conserving 
and CP violating pion-nucleon coupling constants respectively.
Here, $i=0,1,\text{and } 2$ represent isoscalar, isovector and isotensor 
respectively. Considering $\bar{g}_{\pi NN}^{(1)}$ as the most dominant
\begin{equation}
  \bar{g}_{\pi NN}^{(1)} < 6.4 \times 10^{-13}.
\label{gpinn}
\end{equation}
The coupling constant $\bar{g}_{\pi NN}^{(1)}$ is related to the chromo-EDMs of
quarks \cite{pospelov-02}, from the above result
\begin{equation}
   (\widetilde{d}_u - \widetilde{d}_d) < 3.2 \times 10^{-27} e\text{ cm}.
\end{equation}
Next, consider the maximum contribution to NSM arising from 
$\bar{g}_{\pi NN}^{(0)}$, then
\begin{equation}
  \bar{g}_{\pi NN}^{(0)} < 4.5 \times 10^{-12}.
\end{equation}
Since, $\bar{g}_{\pi NN}^{(0)} = 0.027 \theta_{\rm QCD}$ \cite{crewter-79}, 
we get the bound
\begin{equation}
  \theta_{\rm QCD} < 1.7 \times 10^{-10}.
\end{equation}
The value we have obtained for the NSM is likely to give the most stringent
bounds for supersymmetric CP violating phases \cite{pospelov,olive,barr1993}.
In addition, from our results and the experimental data, it is also possible
to set improved limits on the specific CP violating parameters predicted by 
various extensions of the Standard Model, $\epsilon_q^{\rm SUSY}$, 
$\epsilon^{\rm Higgs}$, $x^{\rm LR}$ \cite{romalis01}.

   In conclusion, we have developed a unique relativistic coupled-cluster based 
many-body method that takes into account the physical effects arising from the 
interplay of two very different kinds of fundamental interactions--the  
CP conserving electron-electron Coulomb and CP violating electron-nucleus 
interactions. The results obtained for the EDM of $^{199}$Hg  by the 
application of this method and the latest experiment on this atom 
\cite{griffith-09} yield the most accurate limits to date on some important 
CP violating parameters. The electron correlation effects play a critical 
role in improving the existing limit on these parameters. These limits 
constrain the possible extensions to the Standard Model, thereby enhancing 
our current knowledge of the intriguing phenomenon of CP violation.


    Computations of the results presented in the paper were performed using 
the computing facilities of Center for computational material science, JNCASR, 
Bangalore. Parts of the code used in our computations were written by 
R. K. Chaudhuri. We (BPD and DA) thank the Director and staff of INT, 
University of Washington, Seattle for hospitality during our visit there in 
2008, and  W. Haxton, N. Fortson and B. Heckel for helpful discussions.



\begin{thebibliography}{99}
  \bibitem{landau}
     L.~D.~Landau,
     Sov. Phys. JETP {\bf 5}, 336 (1957).
  \bibitem{luders} 
     G.~Luders,
     Ann. Phys. (N.Y.) {\bf 281}, 1004-1018 (2000).
  \bibitem{christenson-64}
     J.~H.~Christenson,  J.~W.~Cronin, V.~L.~Fitch and R.~Turlay,
     Phys. Rev. Lett. {\bf 13}, 138 (1964).
  \bibitem{abe-01}
     K.~Abe, et al. 
     Phys. Rev. Lett. {\bf 87}, 091802 (2001).
  \bibitem{aubert-01}
     B.~Aubert, et al.
     Phys. Rev. Lett. {\bf 87}, 091801 (2001).
  \bibitem{ibrahim}
     T.~Ibrahim and P.~Nath,
     Rev. Mod. Phys. {\bf 80}, 577 (2008) and references therein.
  \bibitem{dine-04}
     M.~Dine, and A.~Kusenko,
     Rev. Mod. Phys. {\bf 76}, 1 (2004).
  \bibitem{gavela}
     M.~B.~Gavela, P.~Hernandez, J.~Orloff, O.~Pene and C.~Quimbay,
     Nucl. Phys. B {\bf 430}, 382 (1994).
  \bibitem{pospelov}
     M.~Pospelov, and A.~Ritz,
     Ann. Phys. (N.Y.) {\bf 318}, 119 (2005).
  \bibitem{barr1993} 
     S.~M.~Barr,
     Int. J. Mod. Phys. A, {\bf 8}, 209 (1993).
  \bibitem{griffith-09}
     W.~C.~Griffith, M.~D.~Swallows, T.~H.~Loftus, M.~V.~Romalis, 
     B.~R.~Heckel, and E.~N.~Fortson, 
     arXiv:0901.2328 , 2009.
  \bibitem{exp-para}
     J.~M.~Amini, C.~T.Jr. ~Munger and  H.~Gould,
     Phys. Rev. A, {\bf 75}, 063416 (2007).
  \bibitem{barr1992}
     S.~M.~Barr,
     Phys. Rev. D {\bf 45}, 4148 (1992).
  \bibitem{lindgren}
     I.~Lindgren and J.~Morrison,
     {\it Atomic Many-Body Theory},
     Springer-Verlag (1982), Berlin;New York. 
  \bibitem{bartlett-07}
     T.~J.~Bartlett and M.~Musial,
     Rev. Mod. Phys. {\bf 79}, 291 (2007).
  \bibitem{flambaum-02}
     V.~V.~Flambaum and J.~S.~M.~Ginges,
     Phys. Rev. A {\bf 65}, 032113 (2002).
  \bibitem{mohanti}
     A.~K.~Mohanty and E.~Clementi,
     Chem. Phy. Lett., {\bf 157}, 348 (1989).
  \bibitem{chaudhuri-99}
     R.~K.~Chaudhuri, P.~K.~Panda and B.~P.~Das,
     Phys. Rev. A {\bf 59}, 1187 (1999).
  \bibitem{miller} 
     T.~M. Miller and B.~Bederson, 
     Adv. At. Mol. Phys. {\bf 13}, 1(1977).
  \bibitem{rad}
      A.~A.~Radtsig and B.~M.~Smirnov, 
      {\em References Data on Atoms, Molecules, and ions} 
      Springer, Berlin, (1985).
  \bibitem{goebel}
       D.~Goebel and U.~Hohm
       J. Phys. Chem, {\bf 100}, 7710 (1996).
  \bibitem{dzuba-02}
     V.~A.~Dzuba, V.~V.~Flambaum and J.~S.~M.~Ginges,
     Phys. Rev. A {\bf 66}, 012111 (2002).
  \bibitem{latha-08}
     K.~V.~P.~Latha, D.~Angom, R.~K.~Chaudhuri and B.~P.~Das,
     J. Phys. B {\bf 41}, 035005 (2008).
  \bibitem{martensson}
     A.~Martensson-Pendrill,
     Phys. Rev. Lett. {\bf 54}, 11 (1985).
  \bibitem{jesus-05}
     J.~H.~de~Jesus and J.~Engel,
     Phys. Rev. C {\bf 72}, 045503 ( 2005).
  \bibitem{pospelov-02}
     M.~Pospelov,
     Phys. Lett. B {\bf 530}, 123 (2002).
  \bibitem{crewter-79}
     R.~J.~Crewther, P.~Di Vecchiaa, G.~Venezianoa and E.~Witten,
     Phys. Lett. B {\bf 88}, 123 (1979).
  \bibitem{olive} 
     K.~A.~Olive, M.~Pospelov, A.~Ritz, and Y.~Santoso,
     Phys. Rev. D {\bf 72}, 075001 (2005).
  \bibitem{romalis01}
     M.~V.~Romalis, W.~C.~Griffith, J.~P.~Jacobs and E.~N.~Fortson,
     Phys. Rev. Lett. {\bf 86}, 2505-2508 (2001).
\end{thebibliography}
\end{document}